# Reinforcement Learning based Transmission Range Control in Software Defined Wireless Sensor Networks with Moving Sensor

*Anuradha Banerjee[1], Abu Sufian[2], Ali Safaa Sadiq[3] and Seyedali Mirjalili[4]*

[1]Department of Computer Application, Kalyani Government Engineering College, Kalyani, India, e-mail*:* anuradhabanerjee@acm.org
[2]Department of Computer Science, University of Gour Banga, Malda, India, e-mail: sufian@ugb.ac.in
[3]Wolverhampton Cyber Research Institute, School of Mathematics and Computer Science, University of Wolverhampton, Wulfruna Street Wolverhampton, WV1 1LY, UK, email: Ali.Sadiq@wlv.ac.uk
[4]Centre for Artificial Intelligence Research and Optimisation, Torrens University Australia, Fortitude Valley, Brisbane, QLD, 4006, Australia e-mail: ali.mirjalili@laureate.edu.au

**Abstract:** Routing in Software-Defined Wireless sensor networks (SD-WSNs) can be either single or multi-hop, whereas the network is either static or dynamic. In static SD-WSN, the selection of the optimum route from source to destination is accomplished by the SDN controller(s). On the other hand, if moving sensors are there, then SDN controllers of zones cannot handle route discovery sessions by themselves; they can only store information about the most recent zone state. Moving sensors find lots of robotics applications where robots continue to move from one room to another to sensing the environment. A huge amount of energy can be saved in these networks if transmission range control is applied. The multiple power levels exist in each node, and each of these levels takes possible actions after a potential sender node decides to transmit/forward a message. Based on each such activity, the next states of the concerned sender node and the communication session are re-determined while the router receives a reward. The Epsilon-greedy algorithm is applied in this study to decide the optimum power level in the next iteration. It is determined anew depending upon the present network scenario. Simulation results show that our proposed work leads the network to equilibrium by reducing energy consumption and maintaining network throughput.

## 1. Introduction

Software-Defined Wireless sensor networks (SD-WSNs) are an essential component of internet-of-things or IoT. Routing in SD-WSNs is wireless, and it can be either single or multi-hop. The network is either static or dynamic [1, 2]. In static SD-WSN, the selection of the optimum route from source to destination is accomplished by the SDN controller(s) [3]. On the other hand, if moving sensors are there, SDN controllers of zones cannot handle route discovery sessions by themselves. They can only store information about the most recent zone state, which is broadcast throughout the zone after the zone state changes significantly; that is, most nodes go from one neighborhood to another. Static and moving sensors have got their real-life applications [4]. Moving sensors find many robotics applications where robots continue to move from one room to another, sensing the environment. A considerable amount of energy can be saved in these networks if transmission range control is applied. Transmission range control is a technique to reduce the transmission range of senders in each hop depending upon its distance from the intended receiver [5]. The provision of multiple power levels exists in each node, and each of these levels is possible actions after a potential sender node decides to transmit/ forward a message. Based on each such activity, the next states of the concerned sender node as well as the communication session are re-determined while the router receives a reward [6].

To decide the optimum power level in the next iteration, Epsilon ($\sigma$)-greedy algorithm [7] is applied in this article. The value of s is not constant here. It is determined anew depending upon the present network scenario. Simulation results show that our current work, Reinforcement Learning based Transmission Range Control (RL-TRC), leads the network to equilibrium by reducing energy consumption and maintaining network throughput.

### 1.1. Contributions of the present article are as follows:

i) In RL-TRC we have investigated reinforcement-based learning (RL) for transmission range control depending upon both current node state and session state. The importance of the session state follows from the fact that each channel has got its own characteristics in terms of fading, interference, and node trajectory. We have particularly considered moving sensors in the SD-WSN environment.

ii) The entire network is modeled as a multi-agent system where each node consults the environment before taking a transmission range control decision. The learning procedure is entirely decentralized, based on a partially observable Markov Decision Process. Since the observations have Markovian property, remembering only the most recent value of observed parameters is sufficient. One need not take care of the previous values.

iii) Information about the session state is collected from data or control messages from the source of the session. Source propagates that how much energy of routers has been wasted and how much has been appropriately utilized. Also, quality time spent information appears there. Time duration is considered to be quality time if it does not yield an unsuccessful transmission. The current zone's energy and time scenario is also taken care of while deciding a transmission range. Suppose the current location or the current session has already suffered from vast wastage of energy (as well as time). In that case, little weight is given on transmission range control based on the destination's current location. The urge for a bit of energy-saving might incur a huge cost.

iv) An RL-TRC module is designed which can be embedded in any routing protocol. It will boost the protocol's performance in terms of a significant reduction in energy consumption and delay, improving network throughput.

v) Unlike any other related works, RL-TRC is particularly suitable for moving sensors because based on previous communication session experience, it can estimate velocities of successors of a node, approximate attenuation per unit distance considering multiple power levels of nodes, and longevity of links. If a link breaks much before it should, then the link is not regarded as reliable, and RL-TRC does not prescribe shrinking in power level for those. Simulation results show that RL-TRC is much more accurate in predicting optimum transmission range in various situations, producing significantly improved performance in terms of network throughput, message cost, and percentage of alive nodes, energy consumption, and the average delay in packet delivery.

### 1.2 Organization of the article
The introduction in section 1, the technical background in section 2, and related works in section 3. Details model in section 4, simulation studies in section 5 whereas section 6 concludes the article.

## 2. Technical Background

### 2.1. Transmission Range Control for Energy Efficiency

These days' researchers have developed a great interest in Internet-of-Things or IoT. Wireless sensor networks or WSNs form a big part of it [8]. Among various energy-saving techniques in WSN, sleeping strategies, transmission range control, and topology control are mention-worthy. Sleeping strategies instruct an exhausted node to go to sleep for pre-defined / pre-computed time duration, while only lively routers are included in ongoing communication sessions. Topology control techniques are all about monitoring and controlling the locations of sensor nodes. Transmission Range Control (TRC) is widely applied on a per-hop basis where each potential sender restricts its transmission power so that it is just sufficient to reach the intended downlink neighbor. Factors like network density, the energy of nodes involved in the hop, fading of the channel, and node movement trajectory affect it [1, 9, 10]. However, there are a lot of demerits of this approach. Most of the techniques that implement TRC depend on link quality prediction to adapt transmission power to any environment. As per the literature on WSN, routing protocols are broadly classified as proactive and reactive [13]. These rely on empirical studies or analytical models for transmission range control. Empirical studies evaluate the link quality of channels and use it as a metric for transmission range determination. But analytical models have been more successful from a Quality of Service (QoS) perspective because they prepare nodes to adapt to unexpected changes in the network, significantly reducing wasted energy and time. As a direct consequence of this, spatial reuse and contention mitigation are obtained, and ultimately the network converges to a steady-state. Therefore, machine learning (ML) [11] represents an attractive solution where each sensor is modeled as an agent that interacts with the environment. Here, ML decides the next optimum power level using the σ-greedy algorithm [7]. This algorithm picks up the current best option ("greedy") most of the time but sometimes prefers a random option with a small probability; this helps a system get out of local maxima and explore the global one. Therefore, among multiple power levels, RL-TRC does not always prefer the one producing the highest reward for the concerned node.

### 2.2. SD-WSN with moving sensors

SD-WSN is a software-defined wireless sensor network where an SDN controller of each zone keeps track of present zone state and identification numbers as well as residual energies and maximum possible velocities of all nodes present in it [4, 12]. Figure 1 shows a typical picture of SD-WSN with moving sensors. When a node in a zone tries to communicate with some other node, it asks its own zonal controller to know the most recent location of the intended destination. If a recent location of the destination is known to the SDN controller, it checks whether it is possible for the node to cross-boundary of the present zone and enters into some peripheral zone. If it is found that the destination cannot exit periphery of the present zone during the time interval between the last interaction and current time, then the SDN controller informs the corresponding source about the most recent known location of the destination and maximum possible velocity [2]. In response, the source geocasts route-request packets to all nodes within the territory of the pre-estimated broadcast circle expected to embed the present location of the destination. But if the broadcast circle spans multiple zones, then the SDN controller of source interacts with controllers of those neighbor zones so that they can suggest associated peripheral nodes about a much shorter broadcast circle limited within that zone only. The Information about zonal territory is broadcast inside the zone by the SDN controller at regular intervals.

### 2.3. Reinforcement-based Learning for Transmission Range Control

The literature on machine learning can be broadly classified into supervised learning (SL), unsupervised learning (UL), and reinforcement-based learning (RL). The distinguishing characteristic of reinforcement-based learning is that it learns on its own from an unknown environment. Training is done on a batch of data or real-time data. The process of collecting data is observing changes in the environment based on various actions taken.

RL is also suitable for routing in networks [13].

The WSN is modeled as comprising three entities – nodes, live sessions, and the zones. Nodes are agents, while the other two are parts of the environment. Before taking action, each node observes the state of the current zone and the current session, and accordingly, makes its move. WSN abides by the Markovian property; that is, only the last state and action are necessary to compute the probability of arriving at specific states and getting a reward in each iteration. So, it is unnecessary to memorize all the previous values but only the ones that happened in the last event. At the k-th iteration, state of the network is $s_k \in S$ where S is the set of all states. Each node can take an action $a_k \in A$ where A is a set of all actions. Each action influences the environment. Reward rk may be positive or negative. A negative reward is often referred to as punishment. There may be a clash between the reward of an individual entity and the reward of the environment. In the present work RL-TRC, we assume that none of the nodes is selfish. They aim at maximizing the reward of the zone. In section V, we demonstrate the computation of rewards for zone and environment in detail.

### 2.4 σ-greedy algorithm

σ-greedy algorithm is a very popular approach in balancing exploitation exploration trade-offs [7]. The term greedy implies here that mostly network nodes behave the way they should. For example, among multiple power levels, the one that produces the maximum possible reward is chosen as optimal. For example, if σ is set to 0.05, the algorithm exploits the best variant in 95% of iterations, while choice is random in 5% cases. This is quite effective in practice.

The algorithm can be illustrated with a very simple example. Suppose a system with three machines has worked for ten iterations. Machine #1 has delivered in three iterations and earned a sum total of 12 reward points. Machines #2 and #3 worked for 4 and 3 iterations respectively while the corresponding sum of reward points are 10 and 9. So, reward point per iteration of machine #1 is (12/3 = 4) while for machine #2 it is (10/4 = 2.5) and (9/3 = 3) for machine #3. Hence machine #1 has produced the highest throughput and is the best choice at present. For the next, that is, $11^{th}$ run, the machine with highest current average payout is chosen with probability {(1-σ)+(σ/k)} where σ is very small and machines that don't have a highest current payout, are chosen with probability (σ/k) where k denotes the number of

### 3. Related Work

In [1], a dynamic power control scheme is devised using PID (Proportional – Integral – Derivative) controller with a feedback mechanism. Received signal strength indication is used to measure link quality. If link quality is not good enough, then the error signal is input to the controller. The error signal is the difference between transmitted power and received power. The scheme is not reliable in channels with high interference and noise.

An initial beaconing phase is required in [14], where a message is broadcast network-wide at multiple power-levels, and corresponding packet reception rates (PRR) are recorded. Depending upon the PRR value required for individual links, sender nodes in different hops choose power levels. Links that fail to achieve the needed PRR even at maximum transmission level are unreliable and blacklisted. However, this method completely ignores node movement and residual lifetime. Moreover, the beaconing phase is very costly. Overall, zone state is not considered here, which deserves attention because choosing an inappropriate power level leaves an impact on the node itself and the entire zone.

The beaconing phase of [14] is eliminated in on-demand transmission power control (ODTPC) in [15]. It reduces energy consumption and uses RSSI to separate good links from bad ones. It ignores the movement patterns of nodes which are extremely important from the perspective of motion sensors. Adaptive ODTPC or AODTPC [16] incorporates the criterion of channel fading in ODTPC. Unlike

ODTPC, it predicts future values of RSSI using a Kalman filter. Compared to other prediction techniques, the Kalman filter requires a smaller memory and suffers from lesser computational complexity. But this lags behind from the QoS perspective and does not consider node mobility which is a must from moving sensors' point of view.

The scheme proposed in [17], depends on PRR to differentiate good and bad links. Accordingly, topology is controlled in indoor wireless sensor networks. The scheme reduces network contention, especially in dense networks, and improves packet reception rate. In this article, this scheme will be referred to as beacon-PRR.

RSSI is used in [18] to measure link quality. Moreover, an adaptive golden receive power range is defined to preserve the performance from fading vibrations. Two threshold values of signal-to-interference noise ratio (SINR) are defined with which RSSI is compared. If RSSI is above the high threshold, then transmission power is decreased. On the other hand, if RSSI is lower than the low threshold, then transmission power needs to be increased. Otherwise, no change is prescribed. This scheme is subsequently referred to as beacon-RSSI in this article. Frequent injection of beacons at regular intervals is required in [20] too. It uses PRR as a link metric. Although the authors claim that this scheme reduces power consumption and improves the data packet delivery ratio, still beacons eat up a lot of energy. This paper is referred to as beacon-PRR-2. Please note that a low value of the packet reception rate is not always an indication of the fragile link; interference, too, plays a big part.

RLMan [21], an energy management algorithm based on RL, adapts its energy management policy to a time-varying environment, regarding both the harvested and consumption of the node energy. In RLMAN linear function approximations are to make suitable for resource-constrained systems. In another study [22], a green routing algorithm for SDN using a fork and join adaptive particle swarm optimization has proposed. This study had focused on an optimal number of control nodes and optimal clustering of control nodes to maximize the lifetime of the network. In a research work [23], a fusion of artificial intelligence and a mobile agent for energy-efficient traffic control has proposed. In this work, an RL-based artificial agent learns from experience and produces the optimal action in WSNs.

Unlike the above mentioned state-of-the-art works, our proposed scheme RL-TRC mainly concentrates on SD-WSN with moving sensors. RL-TRC emphasizes that generally, nodes in these networks follow their specific movement pattern each day. Based on this, a node can predict whether its immediate successor tends to get out of its radio-circle in a few moments. If this is the situation, then there is little point in controlling the transmission range for energy optimization. Instead, it will be better if communication can be completed before its successor moves out of its radio-range because transmission range control at this position incurs the cost of two retransmissions along with a possible broadcast session – intra-zonal or inter-zonal, whatever. Also, the zonal and session scenario has been considered in RL-TRC. If the current session (the zone) has already suffered from a considerable waste of energy (time), it is better not to take many risks.

## 4. Details of RL-TRC

### 4.1. The Network Model

*i)    Structure*

Let the network in k-th iteration be denoted as an entity $NT_k$. N is the set of all nodes in the network and $SE_k$ is the set of all live sessions in k-th iteration. So,
$NT_k = (N, SE_k)$
Where $N = \{n_a, n_b, n_c, ..., n_p\}$

A live session $s_k$ in iteration k is described by its source $src(s_k)$, destination $dst(s_k)$ and route RUT$(src(s_k), dst(s_k))$ where,
$SE_k = \{ s_k : src(s_k) \in N, dst(s_k) \in N \text{ and } C1\}$ (1)

Where *C1* is a condition as follows:
*C1*: $1 \leq hp\text{-}cnt(RUT(src(s_k), dst(s_k))) \leq H$ (2)

$RUT(n_i, n_j)$ specifies the route from node $n_i$ to $n_j$. $hp\text{-}cnt(R)$ is a function that computes the hop count of route R. H is the maximum possible hop count in the network. The number of hops must be greater than 1 and less than or equal to H where H is the maximum possible hop count in the network.

A Reward of an entity in k-th iteration is the difference between k-th and (k-1)-th iteration. The Reward at 1-th state is dependent upon action at 0-the state. But there is no 0-th iteration because we assumed that the network began its operation with 1-th iteration. Therefore, the reward at 1-th stage of all entities is 0.

The model of a zone and a network are absolutely the same. A zone is a part of the network. It also consists of certain nodes as well as live sessions. As far as message transmission is concerned, each message is divided into multiple packets. After receiving an acknowledgement from destination for the i-th data packet, the source of a communication session transmits the i+1-th data packet.

Various entities in RL-TRC are nodes, sessions, zones, and the network. After each session, a session reward is intimated to the SDN controller by source or appropriate incoming peripheral of the communication session. The structure is shown in fig. 1. It shows a network consisting of 6 zones. Possible shapes of zones are circular, polygonal, and elliptic. Certain nodes have downlink neighbors belonging to other zones. These are called peripheral nodes. These nodes remain static while others can move. There is a centralized controller in each zone, called the SDN controller. The SDN controller of a zone computes zone reward based on its nodes and sessions, whereas there is a centralized network controller that communicates with all zones at high and regular intervals, collects rewards of individual zones, and then sums them up to calculate the overall reward of the network.

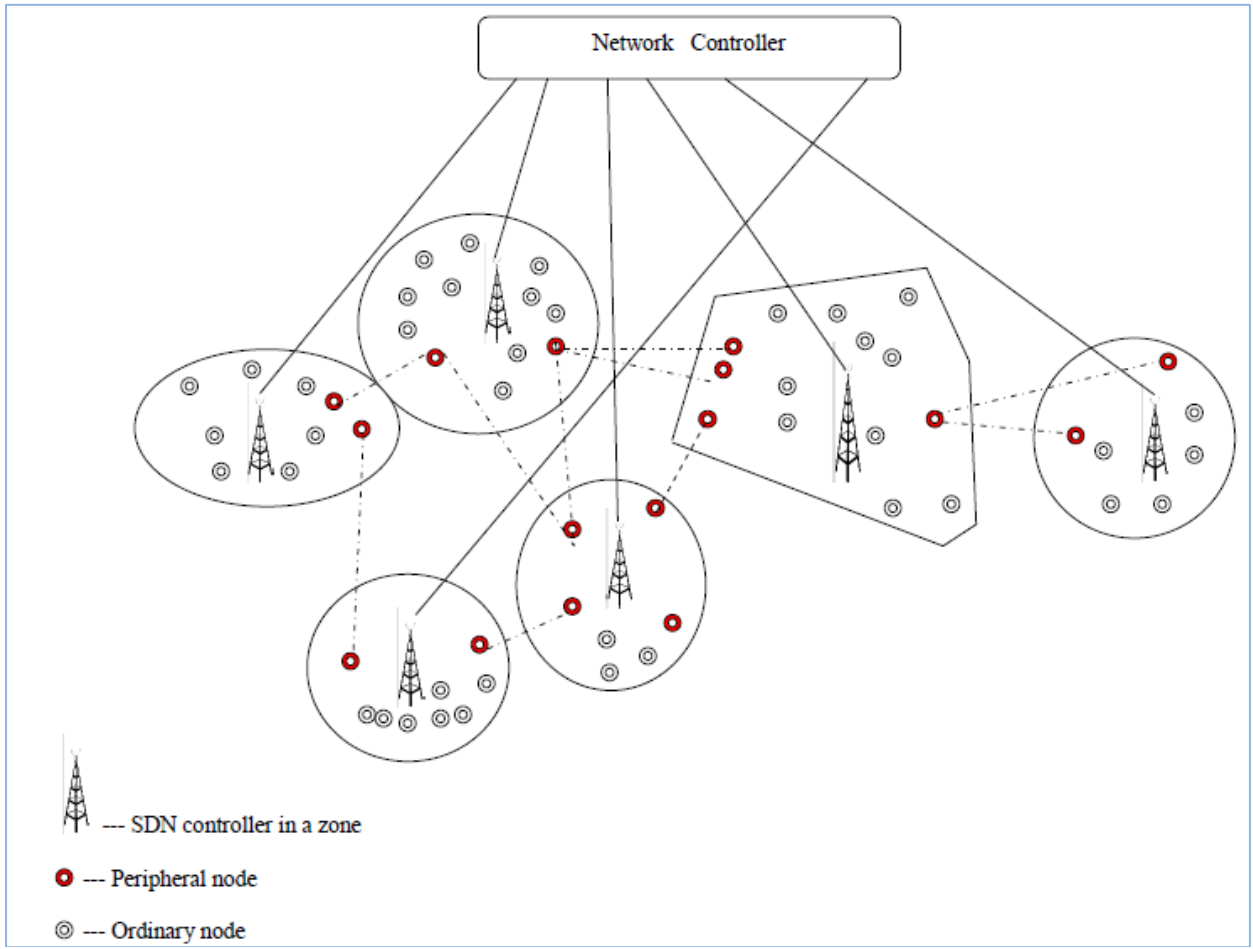

Fig1: Network Framework

Each node $n_i$ computes its own reward as well as rewards of other nodes such as $n_j$ in its eyes, based on the experience of $n_i$ after transmitting a message to $n_j$. In order to store such experiences, a cache memory named communication-cache is maintained. Attributes of the communication cache are as follows:

a) id of the successor node
b) packets transmitted so far
c) packets received so far
d) average strength of received signal
e) average transmission power level
f) recent trend
g) approximate velocity
h) timestamp-begin
i) timestamp-end
j) expected-timestamp-end

The first attribute specifies a unique identifier (say $n_j$) of the node to which $n_i$ has sent the message. The next two attributes are for calculating PRR or packet reception rate. It is equal to (packets received so far/packets transmitted so far). PRR is a positive fraction if acknowledgement of at least one packet has been received by $n_i$ from $n_j$. If no acknowledgement has yet been received by $n_i$ from $n_j$, then PRR of the link from $n_i$ to $n_j$, will be zero. Average strength of received signal is computed as (total strength of all signals received by $n_j$ from $n_i$ so far)/(total number of packets received by $n_j$ from $n_i$ so far). Please note that whenever $n_j$ receives a data packet from $n_i$, it embeds strength of the received signal within acknowledgement packet. This helps $n_i$ to know strength of signal received by $n_j$. Received signal strength or RSS is a measure of signal interference as well as attenuation. A strong value of (RSS/TPL) ratio increases weight or reward of $n_j$, where TPL is average transmission power level. (RSS/TPL) always ranges between 0 and 1. Values close to 1 denote that most of the transmission power of $n_i$, invested to send a message to $n_j$, has been properly utilized.

An idea of a recent trend can be obtained if there is no information to $n_i$ that link with $n_j$ has been broken. If this link is broken, then the recent trend is set to 0. Otherwise, it is 1 or -1. 1 means that $n_j$ is coming closer to $n_i$. -1 denotes that $n_j$ is going further from $n_i$. This is also computed from information in acknowledgement packet. Suppose the timestamps of the last two message packets *PAC1* and *PAC2* are $t_{msg1}$ and $t_{msg2}$ whereas timestamps of corresponding acknowledgements are $t_{ack1}$ and $t_{ack2}$. If $(t_{ack2} - t_{msg2}) > (t_{ack1} - t_{msg1})$ and (average RSS after transmitting *PAC1*) > (average RSS after transmitting *PAC2*), then trend of $n_j$ is going further from $n_i$. On the other hand, if $(t_{ack2} - t_{msg2}) \leq (t_{ack1} - t_{msg1})$ and (average RSS after transmitting *PAC1*) $\leq$ (average RSS after transmitting *PAC2*), then trend of $n_j$ is getting closer to $n_i$. Otherwise, no idea can be obtained about the trend and it continues to be 0.

Approximate velocity can be calculated as follows, assuming that *PAC1* was transmitted with power *PWR1* while corresponding received signal strength *RSS1*. Similarly, *PAC2* was transmitted by $n_i$ with power *PWR2* reciprocated by received strength *RSS2*.

*PAC1* travelled for time $(t_{ack1} - t_{msg1})$ while PAC2 travelled for $(t_{ack2} - t_{msg2})$. Therefore, distance travelled by *PAC1* and *PAC2* are denoted by *dist-PAC1* and *dist-PAC2* and defined in (1) and (2).

$dist\text{-}PAC1 = vs\ (t_{ack1} - t_{msg1})$                  (1)
$dist\text{-}PAC2 = vs\ (t_{ack2} - t_{msg2})$                  (2)

where vs is speed of the wireless signal.

Attenuation suffered by *PAC1* after travelling *dist-PAC1* is (*PWR1-RSS1*) while the same suffered by *PAC2* after travelling *dist-PAC2* is (*PWR2-RSS2*). On an average, signal attenuation per unit distance *SIG-ATN* is formulated in (3).

$$SIG\text{-}ATN = (FF1/\ dist\text{-}PAC1 + FF2/\ dist\text{-}PAC2)/2 \quad (3)$$

Where *FF1= (PWR1-RSS1)*
And *FF2= (PWR2-RSS2)*

*PAC2* suffered from extra *tm(PAC2, PAC1)* amount of time where,

$$tm(PAC2, PAC1) = \{(t_{ack2} - t_{msg2}) > (t_{ack1} - t_{msg1})\} \quad (4)$$

Additional attenuation suffered by *PAC2* is (*FF2 – FF1*) and this is due to travelling extra (*FF2-FF1)/SIG-ATN* distance. $n_j$ travelled *(FF2-FF1)/SIG-ATN* distance in $\{(t_{ack2} - t_{msg2}) > (t_{ack1} - t_{msg1})\}$ time. So, approximate velocity $vel_j(i)$ of $n_i$ based on this trend, is $\{(FF2\text{-}FF1)/SIG\text{-}ATN\} / \{(t_{ack2} - t_{msg2}) > (t_{ack1} - t_{msg1})\}$.

Suppose, $n_i$ now wants to send a new packet *PAC3* to $n_j$ knowing the context that link between them is still alive. Current timestamp is $t_{msg3}$. By this message, $n_j$ is expected to travel the distance *dist-PAC3-est* such that,

$$dist\text{-}PAC3\text{-}est = vel_j(i)\ (t_{msg3} - t_{ack2}) \quad (5)$$

If $\{dist\text{-}PAC3\text{-}est > (2 \times R_i)\}$, then $n_i$ drops the idea of sending messages to it. Otherwise, power level has to be chosen such that it is greater than *P-THRES,* where,

$$P\text{-}THRES > \{(SIG\text{-}ATN \times dist\text{-}PAC3\text{-}est) + min\text{-}rcv(j)\}$$

*min-rcv(j)* is minimum receive power of $n_j$. Basically the power level should be such that even after suffering from estimated signal attenuation, received signal strength should be greater than minimum *RSS* requirement. If $n_i$ has five power levels *PWR1, PWR2, PWR3, PWR4* and *PWR5* such that *PWR1< PWR2< PWR3<PWR4<PWR5* among which only *PWR4* and *PWR5* are greater than *P-THRES*, then available power level range of $n_i$ for $n_j$ will be *{PWR4, PWR5}*. This is termed as "**conditional shrinking of power level set**". Among them *PWR5* is the maximum power level, therefore only *PWR4* is considered as the random option.

*timestamp-begin* is the timestamp at which link from $n_i$ to $n_j$ was established this time. If expected timestamp of link breakage is denoted as $t_{end}$, then,

$$R_i = vel_j(i)\ (t_{end} - t_{ack2})/2 \quad (6)$$

$$t_{end} = 2R_i / vel_j(i) + t_{ack2} \quad (7)$$

If actually the link breaks before $t_{end}$, then the link is not considered reliable. These kinds of links are generally avoided and if at all tried, then only maximum power level is utilized by $n_i$. Otherwise, $\sigma$-greedy is tried with available power level options.

ii) Route-discovery, Route-Selection, and Route-breakage

## 4.2. Route-discovery

When a source node $n_s$ wants to communicate with a destination $n_d$, it broadcasts route-request within the territory of its zone. If a router receives this route-request, it broadcasts the message to all of its downlink neighbors sharing the same zone. As soon as the route-request reaches $n_d$, $n_d$ selects an optimum path as per the underlying protocol. In this article, we have considered three protocols LEACH, SPIN, and TEEN which are considered as standard in the context of wireless sensor networks. These protocols are discussed in detail in under the topic "Route-Selection".

If $n_d$ is found within the zone of $n_s$ only, then it is an intra-zonal phenomenon. Otherwise, multiple zones are involved. SDN controller of the source instructs its peripheral nodes (please note that peripheral nodes are static) to forward it to neighboring zones so that route-requests are circulated to other zones. This is called the inter-zonal broadcast. In the best case, route-requests are confined to the source zone whereas in the worst case, they are flooded throughout the entire network, that is, all zones in the network.

### 4.3. Route-selection

LEACH, SPIN, TEEN, etc. are extremely important from the perspective of non-software defined energy efficiency in sensor networks. The idea of LEACH protocol is to organize the sensor nodes into clusters, where each cluster has one CH that acts as a router to the base station. However, the LEACH protocol has problems, one of those problems is a random selection of cluster head or CHs [24, 25, 26]. This process does not consider the location of sensor nodes in the wireless sensor network, and hence the sensor nodes may be very far from their CH, causing them to consume more energy to communicate with the CH [24, 36, 43]. The role of cluster head is rotated in order to balance the energy consumption among multiple nodes. The optimum route chosen for communication is the shortest one computed using Dijkstra's shortest path algorithm [36]. But the shortest route is not always the most energy-efficient option.

SPIN [29] is another mention-worthy energy-efficient routing protocol which is a modification of classic flooding. In classical flooding the information is forwarded on every outgoing link of a node. This drains out the battery of a huge number of nodes in the sensor network. SPIN was developed to overcome this drawback. It is an adaptive routing protocol, which transmits the information first by negotiating. It proposes the use of metadata of actual data being sent. Metadata contains a description of the actual message the node wants to send. Actual data will be transmitted only if the node wishes to receive it, that is, similar to be keen to watch a movie after viewing its trailer. In this context, we humbly state that SPIN requires the broadcasting of meta-data at least. However, an important aspect of energy-efficiency in TEEN arises from the fact that it does not require broadcasting to establish routes.

TEEN (Threshold-sensitive Energy Efficient sensor Networks) [28] protocol was proposed for time-critical applications. Here sensor nodes sense the medium continuously, but data transmission is done less frequently. A cluster head sensor sends its members a Hard Threshold (HT), which is the threshold value of the sensed attribute and a Soft Threshold (ST), which is a small change in the value of the sensed attribute that triggers the node to switch on its transmitter and transmit. The main drawback of this scheme is that, if thresholds are not received, the nodes will never communicate and the user will not get any data from the network at all. Also, it has the complexity associated with forming clusters at multiple levels and the method of implementing threshold-based functions [28].

### 4.4. Route-breakage

If a route from $n_i$ to $n_j$ breaks when both are routers, then $n_i$ accordingly modifies weight or reward of $n_j$ in its eyes and sends a link breakage message to source $n_s$ of the communication session. Receives this message, $n_s$ modifies rewards of the session, initiates a broadcast which may be intra-zone or inter-zone. Also, $n_s$ informs the SDN controller about this broadcast so that it can update zone reward.

## A. Reward of A Node

For any arbitrary node $n_j$ in the network, let $lvl(j)$ number of power levels are there – $p_1, p_2, p_3, ..., p_{lvl(j)}$. Any action $A_j(k)$ of $n_j$ in k-th iteration is formulated as:

$$A_j(k) \in \{p_m : 1 \leq m \leq lvl(j)\} \qquad (8)$$

Reward $r_j(k)$ of a node $n_j$ till iteration $k$, is formulated as a recursive definition in (9).

$$r_j(1) = 0$$
$$r_j(k) = p_{lvl(j)} - A_j(k-1) + r_j(k-1) \qquad (9)$$

Reward $rd_u(j,k)$ of a successor $n_u$ of $n_j$ in k-th iteration of the network is computed based on the following rules.

**Rule-1**

If $n_j$ received an acknowledgement from $n_u$ in (k-1)-th iteration then,

$$rd_u(j,k) = \begin{cases} (F\text{-}PRR_u(j,k) \times F\text{-}RSS_u(j,k))^{0.5} & \text{if recent trend} = 0 \\ (F\text{-}PRR_u(j,k) \times F\text{-}RSS_u(j,k))^{0.25} & \text{if recent trend} = 1 \\ (F\text{-}PRR_u(j,k) \times F\text{-}RSS_u(j,k))^{0.75} & \text{if recent trend} = -1 \end{cases}$$

where,

$$F\text{-}PRR_u(j,k) = (1+ PRR_u(j,k))/2 \qquad (10)$$

$$F\text{-}RSS_u(j,k) = (1+RSS_u(j,k))/(2 \times TPL_j(u,k)) \qquad (11)$$

Rule-1 is based on the fact that the reward of a node $n_u$ in the eyes of $n_j$, will improve in k-th iteration if $n_j$ has a nice communication experience with $n_u$ in (k-1)-th iteration. $PRR_u(j,k)$ is packet reception rate of the link from $n_j$ to $n_u$ till k-th iteration. Similarly, $RSS_u(j,k)$ and $TPL_j(u,k)$ denote received signal strength and transmission power level of the same link till k-th iteration. If recent trend is not known, then we go on neutral and hence, 0.5 is raised to the power of $(F\text{-}PRR_u(j,k) \times F\text{-}RSS_u(j,k))$. 0.25 is raised to $(F\text{-}PRR_u(j,k) \times F\text{-}RSS_u(j,k))$ if $n_u$ is expected to come close or *trend = 1*. On the other hand, if *trend = -1*, then $(F\text{-}PRR_u(j,k) \times F\text{-}RSS_u(j,k))^{0.75}$ is assigned to $rd_u(j,k)$. Since, $(F\text{-}PRR_u(j,k) \times F\text{-}RSS_u(j,k))$ is a fraction, $(F\text{-}PRR_u(j,k) \times F\text{-}RSS_u(j,k))^{0.75} < (F\text{-}PRR_u(j,k) \times F\text{-}RSS_u(j,k))^{0.5} < (F\text{-}PRR_u(j,k) \times F\text{-}RSS_u(j,k))^{0.25}$. That is, nodes gain weight by coming closer to their predecessor in the hop. But an overall good weight or reward requires a good *PRR* and *(RSS/TPL)* too.

**Rule-2**

If $n_j$ was expecting to receive an acknowledgement from $n_u$ in (k-1)-th iteration but it did not receive that, then *nj* will check whether a maximum number of attempts has already been utilized. Nothing will be done if more attempts are left. But if all chances are exhausted, then cost of broadcasting route-request in the current zone will be deducted from the nodes reward. So,

$$rd_u(j,k) = \begin{cases} rd_u(j,k-1) - broad\text{-}cost(Z) & \text{if } turn(j,k) > mx\text{-}atmpt \\ rd_u(j,k-1) & \text{otherwise} \end{cases} \qquad (12)$$

The above definition is based on the concept that when the network starts functioning, i.e. the first iteration begins, the reward of each node is 0. Until and unless any action is taken, there is no possibility of gain. Gain $r_j(2)$ in 2$^{nd}$ iteration depends on the action taken in 1$^{st}$ iteration. Action is nothing but the chosen power level. The power level chosen by $n_j$ at 1-th iteration is $A_j(1)$. Therefore, gain of $n_j$ is the power saved by it, that is, the difference between the maximum and chosen power level which is mathematically represented by $(p_{lvl(j)} - A_j(1))$.

$turn(j,k)$ specifies the number of turns availed by $n_j$ in iteration k. For example, if the current data packet is being transmitted for the first time, then $turn(j,k) = 1$. If acknowledgement is not received in first turn, same data packet is sent for the second turn. Similarly, if acknowledgement is not received in second turn, that packet is sent for the third turn. If current data packet is transmitted for second or third time, $turn(j,k)$ will take the value 2 or 3, respectively. $mx$-$atmpt$ is the maximum number of times same data packet can be transmitted. Generally, value of $mx$-$atmpt$ is 3, that is, a maximum of two retransmissions are allowed. If in (k-1) th iteration, a packet was transmitted by $n_j$ for the first time with corresponding power level or action $A_j(k-1)$, gain in k-th iteration increases by $(p_{lvl(j)} - A_j(k-1))$ from reward $r_j(k-1)$ of node $n_j$ till $(k-1)th$ iteration of the network.

If the acknowledgement is not received even after the third turn, then route-request has to be broadcast by source (if the source is inside the current zone) or the incoming peripheral with least id (after being instructed by $n_j$), at least throughout the current zone, to discover a new route to the same destination. This involves some cost which is termed as broadcast cost.

Below we compute average broadcast cost $broad$-$cost(Z)$ in a zone Z, as the average of minimum and maximum broadcast cost in a zone. Please note that minimum broadcast cost in a zone corresponds to minimum hop count and minimum hop count is associated with maximum progress in each hop. Following Lemmas 1 and 2 compute minimum value of hop count and the minimum broadcast cost in a zone, respectively.

**Lemma 1:** Minimum hop count in a zone Z is $\{\theta(Z)(2\phi(Z)+1)/( 2\phi(Z)\ av$-$rad(Z))\}$ where $\theta(Z)$ is the maximum possible distance between any two nodes in the zone Z, $\phi(Z)$ is an average number of downlink neighbors whereas $av$-$rad(Z)$ denotes average radio-range in Z.

Proof: Let, set of downlink neighbors of a node at time t be denoted as $D_i(t)$. For any node $n_u \in D_i(t)$, current distance between $n_i$ and $n_u$ is denoted by $a$ and the angle between them, by $b$. We have omitted subscripts here for simplicity. Here $0 \leq a \leq R_i$ and $-\pi/2 \leq b \leq \pi/2$. Please consider figure 2.

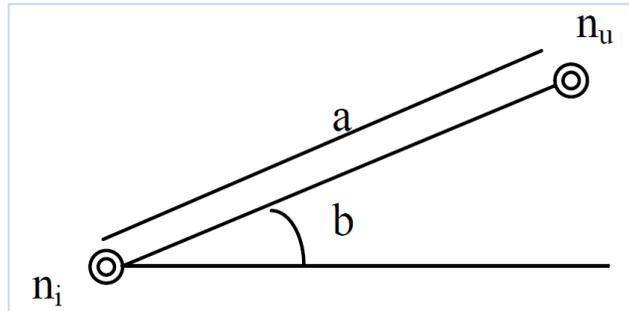

**Fig 2: $n_u$ is at distance $a$ from $n_i$ and makes an angle $b$ with it**

The probability distribution function $F(a,b)$ of distances between $n_i$ and present set of downlink neighbors including the angle, is,

$F(a,b) = 2a / (\pi R_i^2)$ (13)

Integrating F(a,b) over b gives pdf of a as f(a) where,

$$f(a) = \int_{-\pi/2}^{\pi/2} 2a/(\pi R_i^2)\, db \qquad (14)$$

$$f(a) = 2a/R_i^2 \qquad (15)$$

Based on (15), probability distribution function $f(a_{max})$ of the maximum distance between $n_i$ and its present set of downlink neighbors, is modeled below.

$$f(a_{max}) = |D_i(t)|\, (g(a))^{(|D_i(t)|-1)}\, f(a) \qquad (16)$$

where $g(a)$ is the cumulative distribution function of $a$. It is formulated in (17).

$$g(a) = \int_{-\infty}^{a} (2y/R_i^2)\, dy \qquad (17)$$

i.e.,

$$g(a) = \int_{0}^{a} (2y/R_i^2)\, dy = a^2/R_i^2 \qquad (18)$$

Putting this value of $g(a)$ in (16), we get,

$$f(a_{max}) = 2|D_i(t)|\, a^{2(|D_i(t)|)-1}/R_i^{2|D_i(t)|} \qquad (19)$$

Expectation $E(a_{max})$ is computed in (20).

$$E(a_{max}) = \int_{0}^{R_i} a\, f(a_{max})\, da \qquad (20)$$

i.e., $E(a_{max}) = 2|D_i(t)| \int_{0}^{R_i} a^{2(|D_i(t)|)}/R_i^{2|D_i(t)|}\, da$

i.e., $E(a_{max}) = 2|D_i(t)|\, [a^{2(|D_i(t)|)+1}/R_i^{2|D_i(t)|}]_{0}^{R_i}/(2|D_i(t)|+1)$

i.e., $E(a_{max}) = 2|D_i(t)|\, [R_i^{2(|D_i(t)|)+1}/R_i^{2|D_i(t)|}]/(2|D_i(t)|+1)$

So, $E(a_{max}) = 2|D_i(t)|\, R_i/(2|D_i(t)|+1) \qquad (21)$

On an average, for any arbitrary zone $Z$, if $\phi(Z)$ and *av-rad(Z)* indicate average number of downlink neighbors and average radio-range of a node, then average maximum progress *prg(Z)* per hop to the destination in zone $Z$, is in equation (22)

$prg(Z) = 2\phi(Z) \, av\text{-}rad(Z) /( 2\phi(Z)+1)$ (22)

Therefore, minimum hop count *h-min(Z)* in zone Z will be $\{\theta(Z)(2\phi(Z)+1)/( 2\phi(Z) \, av\text{-}rad(Z))\}$ where $\theta(Z)$ is maximum possible distance between any two nodes in the zone Z. Maximum hop count, on the other hand, corresponds to minimum progress per hop, which is 1. Therefore, maximum hop count is $\theta(Z)$. So, average value *h-avg* of the hop count is given by,

$h\text{-}avg = \theta(Z)\{1+(2\phi(Z)+1)/( 2\phi(Z) \, av\text{-}rad(Z))\}/2$

Based on this, average broadcast cost broad-cost in zone Z is computed in (23) where ng is average number of downlink neighbors of nodes belonging to zone Z. This value is computed by SDN controller in zone Z.

$broad\text{-}cost(Z) = ng+ng^2+ng^3+\ldots+ng^{h\text{-}avg(Z)}$ (23)
i.e. $broad\text{-}cost(Z) = \{1+ng+ng^2+ng^3+\ldots+ng^{h\text{-}avg(Z)}\}-1$
i.e. $broad\text{-}cost(Z) = (ng^{h\text{-}avg(Z)+1}-1)/(ng-1)-1$

### B. Reward of a Session

Let $sn_k$ be a live session in zone *Z* in k-th iteration. Then energy-wasted till k-th iteration of the a zone is recursively modeled as the summation of energy wasted till (k-1) th iteration of the same zone and energy wasted by transmissions belonging to various sessions live in k-th iteration. Mathematical expression of this appears in (24). Similarly expression of wasted time appears in (25).

$ew_k(Z) = ew_{k-1}(Z)+\sum wst\text{-}enrg(\rho, turn_k(\rho))$ (24)
$\quad\quad\quad \rho \in trans(k, sn_k)$
$\quad\quad sn_k$ is a live session in Z

$et_k(Z) = et_{k-1}(Z)+\sum wst\text{-}tme(\rho, turn_k(\rho))$ (25)
$\quad\quad\quad \rho \in trans(k, sn_k)$
$\quad\quad sn_k$ is a live session in Z

$ew_k(Z)$ and $et_k(Z)$ denotes energy and time wasted till k-th iteration of the zone Z. $trans(k, sn_k)$ is the set of transmissions belonging to session $sn_k$ that took place throughout-th episode. Each $\rho \in trans(k, sn_k)$ is represented as an ordered triplet *(hop-start($\rho$), hop-end($\rho$), $turn_k(\rho)$)*. *hop-start($\rho$)* and *hop-end($\rho$)* denote predecessor and successor nodes in $\rho$-th transmission or hop. $turn_k(\rho)$ is the number of times the same packet has been transmitted from *hop-start($\rho$)* to *hop-end($\rho$)*. Please note that *wst-enrg($\rho$, $turn_k(\rho)$)* and *wst-tme($\rho$, $turn_k(\rho)$)* indicate energy and time wasted in $\rho$-th transmission where $turn_k(\rho)$ is the number of attempts.

If $turn_k(\rho)$ is equal to 1 it means that the current data packet is being transmitted for the first time; therefore no energy and time has been wasted by it. Wastage of energy and time results from unsuccessful transmissions. Hence, if $turn_k(\rho) = 2$, then energy level chosen in 1-th transmission will be treated as wastage, and sum of corresponding transmission time and time to receive an acknowledgement will be treated as wasted time. Although 1-th transmission was a part of (k-1) th episode, still (k-1)th episode cannot distinguish whether the transmission was successful or not. If an acknowledgement is received in k-th episode then transmission in (k-1)th episode was successful; otherwise not and therefore 2-th transmission needs to be initiated in k-th episode for which selection of power level is required.

If $turn_k(\rho) = mx\text{-}atmpt+1$, it means that the current data packet could not be successfully delivered from hop-start($\rho$) to hop-end($\rho$) even after trying maximum allowable number of attempts. Therefore, route-requests need to be broadcast, throughout the current zone Z in case of intra-zone broadcast and throughout some more zones during inter-zone broadcast. Energy and time required by this is considered to be a wastage. Based on this concept, $wst\text{-}enrg(\rho, turn_k(\rho))$ and $wst\text{-}tme(\rho, turn_k(\rho))$ are mathematically modeled below.

If $turn_k(\rho) = 1$,
$wst\text{-}enrg(\rho, turn_k(\rho)) = wst\text{-}tme(\rho, turn_k(\rho)) = 0$ (26)

If $1 < turn_k(\rho) \leq mx\text{-}atmpts$,
$wst\text{-}enrg(\rho, turn_k(\rho)) = act(hop\text{-}start(\rho), turn_k(\rho) - 1)$ (27)
$wst\text{-}tme(\rho, turn_k(\rho)) = \tau_a$ (28)

Here $act(hop\text{-}start(\rho), turn_k(\rho) - 1)$ is the action or power level chosen by the node hop-start($\rho$) in the immediate previous occasion i.e. $(turn_k(\rho) - 1)$-th occasion when hop-start($\rho$) sent the same packet to hop-end($\rho$). $\tau_a$ is the time period for which a node waits after transmitting a message.

If $turn_k(\rho) = mx\text{-}atmpt+1$
$wst\text{-}enrg(\rho, turn_k(\rho)) = f1(\rho,k) + f2(\rho,k) + f3(\rho,k)$ (29)

$f1(\rho,k) = act(hop\text{-}start(\rho), turn_k(\rho) - 1)$
$f2(\rho,k) = \sum_{ZB \in zone\text{-}set} broad\text{-}cost(ZB)$
$f3(\rho,k) = pwr\text{-}invest(src(sn_k), hop\text{-}start(\rho))$

when $turn_k(\rho) = mx\text{-}atmpt+1$, it means that no more retransmissions are possible. Therefore, the previous transmission of the same data packet from node hop-start($\rho$) intended to hop-end($\rho$) is also in vain. Corresponding wasted energy is $f1(\rho,k)$. Route-request packets need to be broadcast, at least inside the zone, to discover another route to the same destination. The cost of that broadcast is considered to be a component of wastage. It is denoted as $f2(\rho,k)$. As mentioned earlier, zone-set is the set of zones involved in the broadcasting of route-requests. Also, the energy invested by source of $sn_k$ (denoted by $sn_k$), along with all routers till the current one (denoted by hop-start($\rho$)), is also considered to be a wastage because of this energy $pwr\text{-}invest(src(sn_k), hop\text{-}start(\rho))$ could not yield in a successful data packet delivery to intended destination. $f3(\rho,k)$ denotes the associated wasted energy.

$wst\text{-}tme(\rho, turn_k(\rho)) = f1'(\rho,k) + f2'(\rho,k) + f3'(\rho,k)$ (30)

$f1'(\rho,k) = \tau_a$
$f2'(\rho,k) = \sum_{ZB \in zone\text{-}set} \theta(ZB) / prg(ZB)$
$f3'(\rho,k) = time\text{-}invest(src(sn_k), hop\text{-}start(\rho))$

$f1'(\rho,k)$ specifies the time duration for which the node hop-start($\rho$) waited to receive an acknowledgement from hop-end($\rho$), after $mx\text{-}atmpt$ number of attempts. $f2'(\rho,k)$ is the time required for broadcast operation. For broadcast operation, $\theta(Z)$ is the maximum distance to be covered and $prg(Z)$ is per hop approximate progress in zone Z. For inter-zone broadcast, the broadcast may be performed in a serial or parallel fashion. For simplicity, here we have assumed broadcasting in serial fashion. Therefore overall time required for broadcasting in multiple zones is the summation of broadcast duration of each zone in zone-set. $f3'(\rho,k)$ is the time invested by source of $sn_k$ (denoted by $sn_k$), along with all routers till the current one (denoted by $hop\text{-}start(\rho)$).

Reward $R(sn_k)$ of a session is is modeled in (31).

$$R(sn_k) = \{- ew_k(Z)\}^{\{1 - 1/(1+etk(Z))\}} \quad (31)$$

The above formulation is based on the factor that reward of a session reduces with increase in wasted energy and time.

### C. Reward af a Zone and the Network

Reward $R1(Z_k)$ of a zone $Z$ in k-th iteration is modeled in (32). This is based on the concept that the reward of a zone is the sum of two factors – summation of rewards of all nodes in $Z$ and summation of rewards of all live sessions in $Z$.

$$R1(Z_k) = \sum_{n_j \in Z} r_j(k) + \sum_{sn_k \text{ is a live session in } Z} R(sn_k) \quad (32)$$

Assuming that the current network $N$ consist of $V$ number of zones $Z1, Z2, Z3, \ldots, ZV$. Then reward $R'(N_k)$ of the network $N$ n iteration $k$, is modeled in (33). Reward of a network is sum of reward of all of its zones.

$$R'(N_k) = \sum_{1 \leq B \leq V} R1(ZB_k) \quad (33)$$

### D. Selection of Power Level: σ-greedy

Each node selects its optimum power level through σ-greedy algorithm. By virtue of σ-greedy algorithm, if σ is very low then, the highest power level is selected; otherwise any one of the random options among "available power levels" is chosen. Available power level is the set of power levels after conditional shrinking. The highest power level becomes mandatory if the reward of the current zone is very low, that is, already the current zone is suffering from huge wastage of energy and time. The effect is coupled with small reward of the entire network. Based on this, the value of σ is computed in (34).

$$\sigma = \begin{cases} 0.001 & \text{if } R1(Z_k) < 0 \\ R1(Z_k) & \text{if } 0 \leq R1(Z_k) < 1 \\ 1 - 1/(1 + R1(Z_k))^{(|R'(Nk)|+1)} & \text{if } C1 \text{ and } C2 \text{ and } C3 = 1 \\ 1 - 1/(1 + R1(Z_k))^{1/(|R'(Nk)|+1)} & \text{if } C1 \text{ and } C2 \text{ and } C31 = 1 \\ \{1 - 1/(1 + R1(Z_k))\}^{1/R'(Nk)} & \text{if } C1 \text{ and } C21 = 1 \end{cases} \quad (34)$$

The conditions *C1, C2, C3, C31, C21* are a follows:
*C1*:: $R1(Z_k) \geq 1$, *C2*:: $R'(N_k) < 0$, *C31*:: $|R'(N_k)| \leq 1$
*C21*:: $R'(N_k) > 1$

### E. Message Forwarding in RL-TRC

The entire process of message forwarding in RL-TRC is shown in fig. 3 in the form of a flowchart. Whenever any message arrives at a node, its place in a message queue is decided by the underlying scheduler. If the message queue is empty at that time, then the new packet is inserted in 0-th position of the queue. Forwarding of packets always begins from 0-th position of the message queue. Before forwarding the message packet in 0-th position, value of σ needs to be calculated in order to determine the suitable transmission power level. This is the first time that data packet is being forwarded by the current node; therefore turn is set to 1. Then the packet is transmitted to the successor. If an acknowledgement is received then it means that earlier transmission was successful. Then in the next iteration, rewards of all entities are calculated anew if state has changed significantly. After that attention is shifted again to message queues where subsequent messages are to be forwarded. On the other hand, if no acknowledgement is received after the first attempt, value of turn is incremented by 1 and again the packet is transmitted to the successor. The cycle continues till all possible transmission attempts are not exhausted. If turn > mx-atmpt then it means the same message packet cannot be transmitted again to the same successor. A new route to the desired destination needs to be discovered. In new iteration, rewards of all entities are calculated if their state has changed significantly and attention is shifted to message queue.

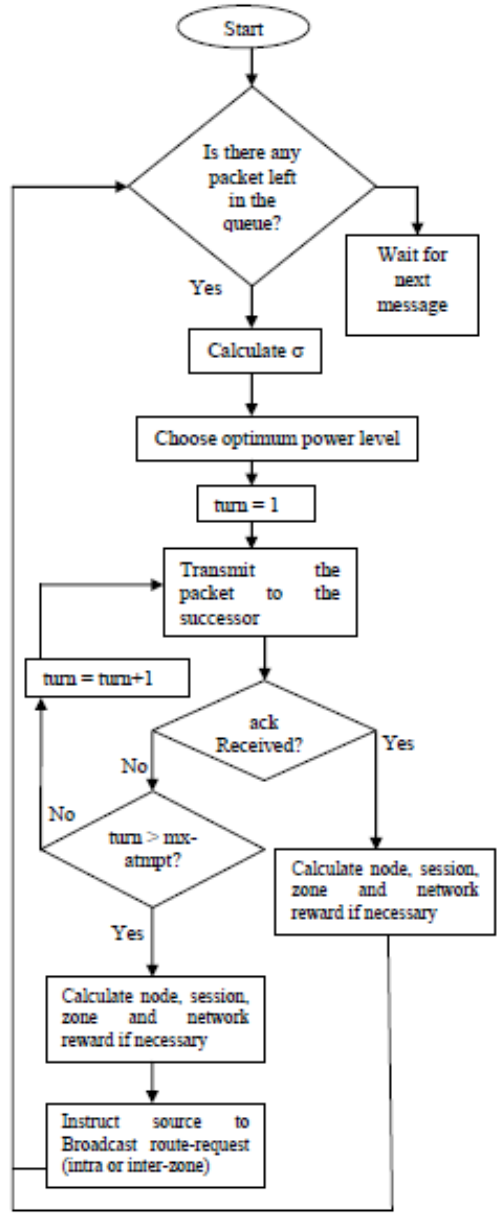

**Fig 3: Forwarding in RL-TRC**

1. **Simulation Experiments and Results**

5.1 **Simulation Environment**

RL-TRC is implemented in the Network Simulator NS2 [30]. The SDWSN is set up with zones of various shapes. The number of zones and the number of nodes in different nodes varies in different simulation runs. The transmitter generates a stream of packets with Poisson distribution and variable inter-arrival time with lower and upper bounds mentioned in table 1. The size of the packets' payload is constant and equal to 50 bytes, over the maximum payload of 123 bytes allowed in the standard. Table 1 shows all the simulation parameters. Ordinary versions of the protocols SPIN, LEACH, and TEEN are compared with ODTPC, beacon-PRR, beacon-RSS, and RL-TRC versions of these protocols.

Table 1: Simulation Parameters

| Parameter | Value |
|---|---|
| Number of zones | 3, 6, 9 and 12 in different runs |
| Minimum number of nodes in each zone | 5 |
| Maximum number of nodes in each zone | 150 |
| Number of nodes | 100,150,200,250,300, 350,400,450,500 |
| Network area | 2000m×2000m |
| Radio-range | 10 m-40m |
| Initial energy of nodes | 20 J – 50 J |
| Minimum number of transmission power level | 1 |
| Minimum number of transmission power level | 25 |
| Minimum packet inter-arrival time | 50 ms |
| Maximum packet inter-arrival time | 200 ms |
| Maximum number of transmission attempts for both message and ack | 4 |
| Communication protocols implemented | LEACH, SPIN, TEEN |
| Mobility model | Random waypoint, random walk and Gaussian |
| Communication standard | IEEE 802.15.4 |
| Competitors of RL-TRC | ODTPC, beacon-PRR, beacon-RSSI |

**5.2. Simulation Metrics**

i) *Overall Message Cost (OMC)* - This is a summation of messages sent by all nodes in the sent by $n_u$ throughout the simulation period.

$$OMC = \sum_{n_u \in N} mg(u)$$

ii) *Energy Consumed (EC)* – This is a summation of energy consumed by all nodes in the network N. It is expressed in joules.

$$EC = \sum_{n_u \in N}(start\text{-}eng(u) - end\text{-}eng(u))$$

iii) *Network Throughput (NTG)* – This denotes the percentage of success in data packet delivery. Assume that total trans-pac number of packets were transmitted throughout the simulation period, among which delv-pac number of packets were successfully delivered to their destinations. Then, network throughput NTG is formulated below.

$$NTG = (dlv\text{-}pac / trans\text{-}pac) \times 100$$

iv) *Average Delay (ADL)* – This is summation of delay faced by all packets in the network, divided by number of packets transmitted. This is denoted as ADL and formulated below.

$$ADL = \sum_{pac \in PCK}\{delv\text{-}tmp(pac)\text{-}gen\text{-}tmp(pac)\} / |PCK|$$

The above formulation is based on the assumption that *PCK* is the set of all packets transmitted throughout the simulation period. *delv-tmp(pac)* and *gen-tmp(pac)* denote timestamp of delivery and generation of the timestamp of the packet *pac*. The difference between them is the delay faced by a packet.

v) *Percentage of alive nodes (PALN)* – This is number of alive nodes (AV) after the end of simulation multiplied by 100 and divided by |N| where N is the set of all nodes in the network.

$$PALN = (AV \times 100) / (|N|)$$

vi) *Average percentage of wasted energy (AWE)* – Let throughout the simulation period, IE amount of energy was invested and UE amount of energy has been successfully utilized. So, the average percentage of wasted energy or AWE is formulated below.

$$AWE = (1-UE/IE) \times 100$$

vii) *Average percentage of wasted time (AWT)* – Let throughout the simulation period, IT amount of energy was invested and the UT amount of energy has been successfully utilized. So, the average percentage of wasted energy or APW is formulated below.

$$AWT = (1-UT/IT) \times 100$$

**5.3 Why RL-TRC is superior?**

In ordinary transmission range control, the first sender node in current hop calculates minimum required transmission power. This is based on the minimum receive power of the receiver node in the current hop and distance between the sender and the receiver. Then among multiple power levels associated with it, the one which is greater than and closest to the minimum required transmission power, is used for message packet transmission. It does not consider the context and communication experience of various other nodes in its zone as well as the scenario of the network. Unlike any other competitor, RL-TRC is particularly suitable for moving sensors because based on previous communication session experience, it can estimate velocities of successors of a node, approximate attenuation per unit distance considering multiple power levels of nodes and longevity of links. If a link breaks much before then it should, then the link is not considered reliable and RL-TRC does not prescribe shrinking in power level for those.

These are the factors that save significant energy and terribly affect network throughput. If the current zone is terribly suffering from scarcity of energy, then it means that a huge amount of energy has already been wasted. Hence, being too much miser to save transmission energy, may wreak havoc in the communication system of the zone (and the network). Retransmissions require additional energy depending on various energy levels chosen in each of them. Also, some valuable time is wasted. After transmitting each message packet, the sender node waits for $\tau_a$ an amount of time to receive an acknowledgment. In general, three transmissions, that is, two retransmissions are allowed in sensor networks. So, after sending a message for the first time, the sender in a hop waits for ($3\tau_a$) amount of time before concluding that the link has been broken. In that case, it sends a route-error message to the source of the incoming peripheral with the least identification number. Route-error message instructs the node to again broadcast route-request messages to discover a new route to the same destination. All these eat up energy in almost every node in the current zone reducing average node lifetime. As a result, some other links break, and more broadcasting sessions of route-request are initiated consuming more energy of nodes in the zone, giving rise to an ominous cycle. The phenomenon prevents many nodes from reaching their destinations. Hence, the data packet delivery ratio decreases.

On the other hand, our present protocol RL-TRC tries to gather experience from most recent records of rewards of the concerned entities. From the earlier record of link quality, the sender in each hop decides the most suitable power level. If the background stage has become unstable because of huge wastage of energy and time, maximum effort should be made to send the message in a single transmission because retransmissions mean unnecessary additional energy consumption and additional waiting time to receive an acknowledgment. If sender in a hop always prefers the energy level that will yield maximum reward to itself (you may call it a selfish move) then it may harm other nodes in the network. The move may result in failure of all retransmissions which, in turn, compels a source or incoming peripheral to broadcast route-request. The situation worsens if route-requests are broadcast across multiple zones. RL-TRC appreciates the fact that the behavior of each node influences the entire network. Hence, nodes should decide upon their transmission range after considering the scenario of the zone as well as network. Therefore, RL-TRC is more rational and prudent in this respect; it does not sacrifice pounds for pennies. Hence, it greatly reduces energy consumption and latency improving the data packet delivery ratio.

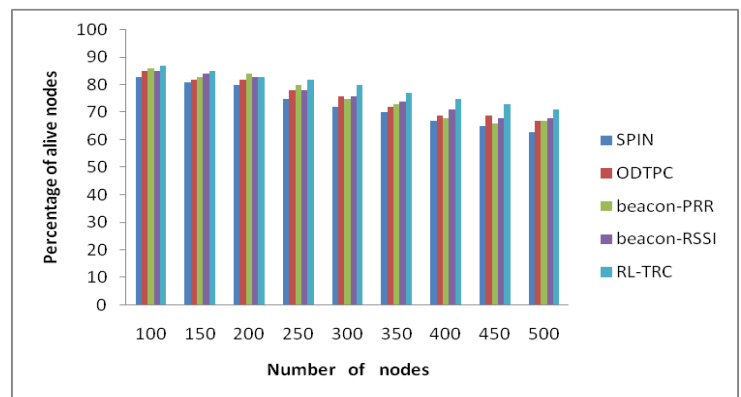

Fig 4. Percentage of alive nodes vs number of nodes (Protocol = SPIN)

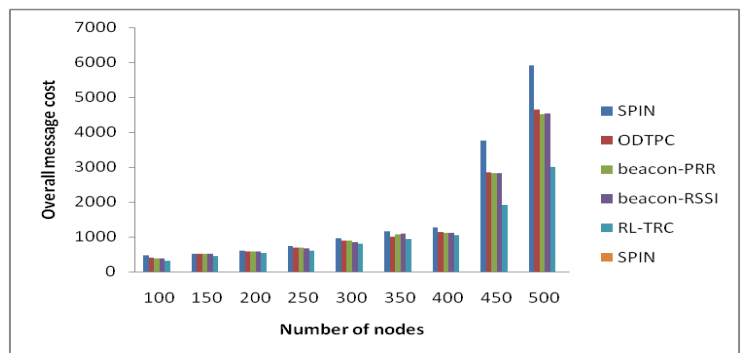
Fig 5. Overall message cost vs number of nodes (Protocol = SPIN)

## 5.4 RL-TRC vs ODTPC:

ODTPC is concerned with the only RSSI of the links and based on this RSSI (received signal strength indication) good links are distinguished from bad ones. But only RSSI is not sufficient to identify the quality of a link. Suppose a node $n_i$ sends 10 packets to $n_j$. Among them, $n_j$ acknowledges only 3 packets but with a high received signal strength report. Therefore, the link from $n_i$ to $n_j$ is considered to be an efficient one, although it is not so. The packet reception rate or PRR has to be combined with RSSI to correctly evaluate a link. RL-TRC combines those two along with link longevity prediction which is based on current movement trend analysis. Successors that are getting closer are far better than those which are going further. This saves a lot of energy and improves network throughput. Also it saves the time and energy of broadcasting route-requests. Hence, the percentage of alive nodes and average packet delay also get improved. RL-TRC embedded versions of protocols (LEACH, SPIN, TEEN, etc.) are far better than other transmission range control techniques embedded versions of those protocols and the improvements are evident from figures 4 to 18.

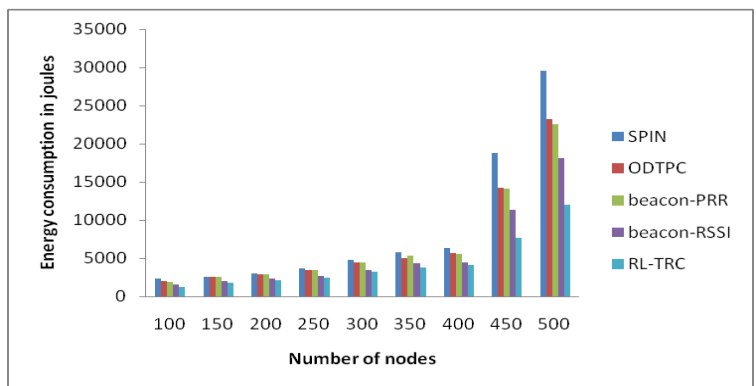
Fig 6. Energy consumption in joules vs number of nodes (Protocol = SPIN)

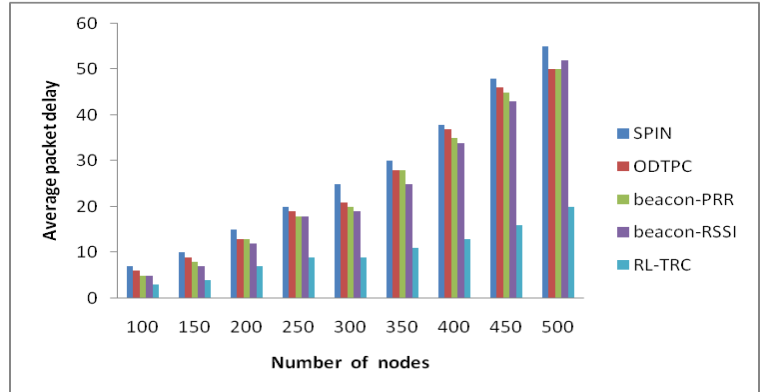

Fig 7. Average packet delay in ms vs number of nodes (Protocol = SPIN)

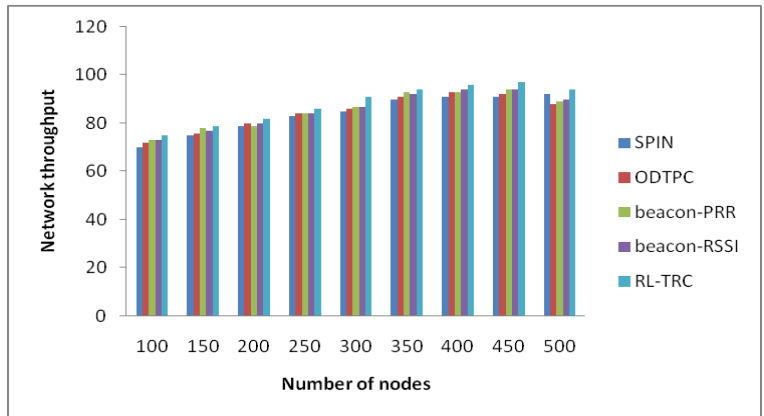

Fig 8. Network throughput vs number of nodes (Protocol = SPIN)

### 5.5 RL-TRC vs beacon-PRR:

Beacon-PRR analyzes a channel based on only PRR. But PRR alone is not only sufficient as we have already mentioned. RL-TRC combines PRR with RSSI and link longevity. If $n_j$ is coming close to $n_i$, then weight of the link from $n_i$ to $n_j$ will increase. On the other hand, if $n_j$ tends to get out of the radio-range of $n_i$, then the weight of the link from $n_i$ to $n_j$ will increase. Based on the estimated position of successor $n_j$, $n_i$ can choose a lower limit of power level so that transmitted signals can overcome the hindrance of attenuation and interference. This conditional shrinking of power level set prevents a node from transmitting signals at power levels that will never reach the intended destination. Hence, RL-TRC also prevents the broadcasting of route-requests resulting due to unsuccessful transmissions that yield from an unsuitable choice of power levels where the transmitted signal gets completely attenuated before reaching the intended moving destination. Saving of energy improves average node lifetime, network throughput, and reduces message cost as well as energy consumption. The improvements are evident in figures 4 to 18.

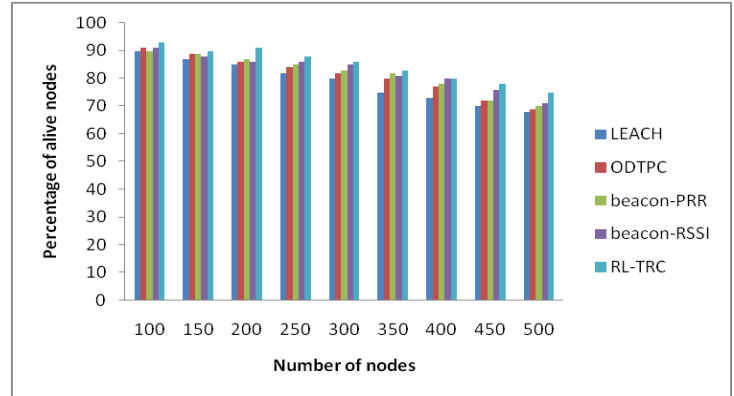
Fig 9. Percentage of alive nodes vs number of nodes (Protocol = LEACH)

### 5.6. RL-TRC vs beacon-RSSI:

Like beacon-PRR, beacon-RSSI also suffers from a huge energy cost due to the transmission of beacons. Beacon-RSSI defines two thresholds – high and low. If RSSI is above the high threshold, the power level is decreased. On the other hand, if RSSI is less than the low threshold, the power level is increased. But, as already mentioned, only RSSI is not sufficient to assess the quality of a link. Moreover, this scheme is not suitable for moving sensors. Often beacon-RSSI wastes energy in attempts to transmit to successors who has already crossed its radio-range. Link lives are often correctly predicted in RL-TRC. It stops nodes from transmitting to neighbors who are not its downlink neighbors anymore. This results in a significant saving in energy as well as node lifetime and network throughput. Also, the average packet delay reduces because time wasted to broadcast route-requests are saved in many cases. Results are evident from figures 4 to 18.

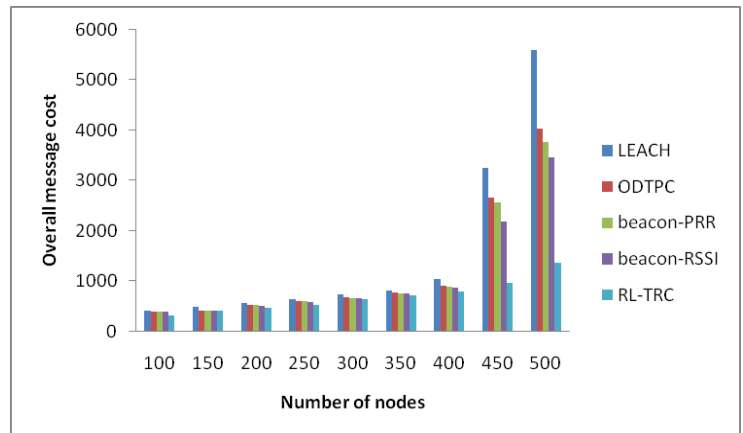
Fig 10. Overall message cost vs number of nodes (Protocol = LEACH)

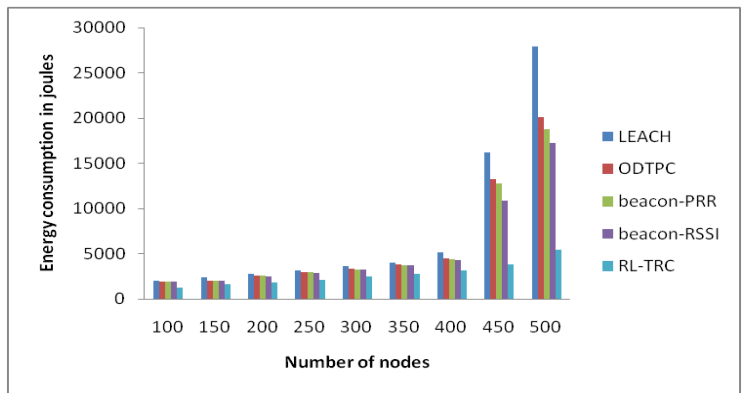
Fig 11. Energy consumption in joules vs number of nodes (Protocol = LEACH)

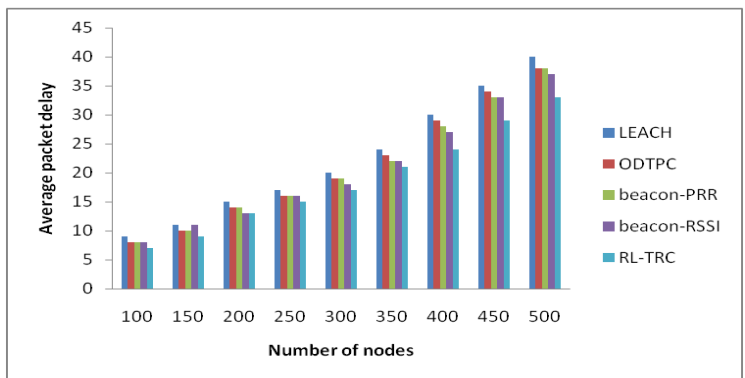
Fig 12. Average packet delay in ms vs number of nodes (Protocol = LEACH)

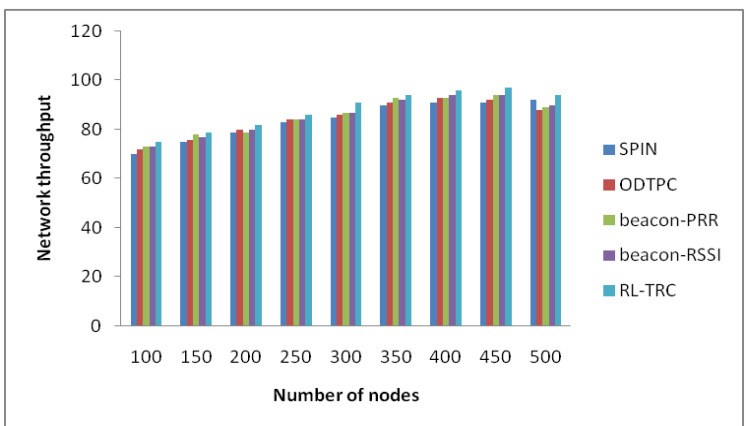
Fig 13. Network throughput vs number of nodes (Protocol = LEACH)

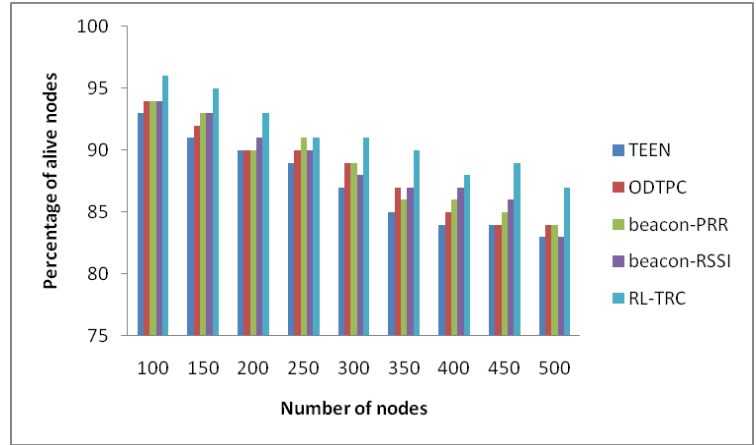
Fig 14. Percentage of alive nodes vs number of nodes (Protocol = TEEN)

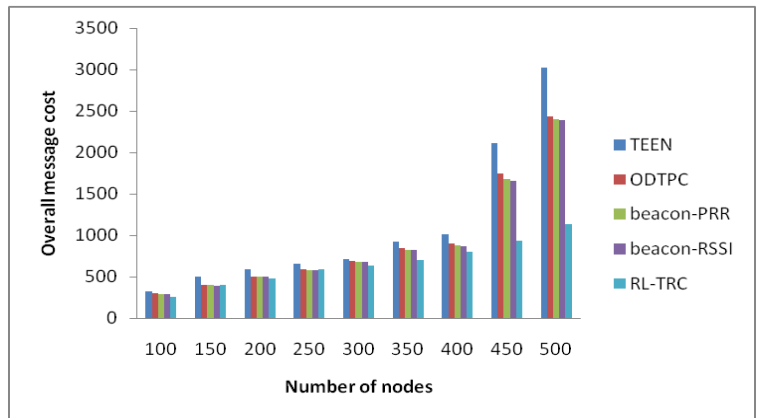
Fig 15. Overall message cost vs number of nodes (Protocol = TEEN)

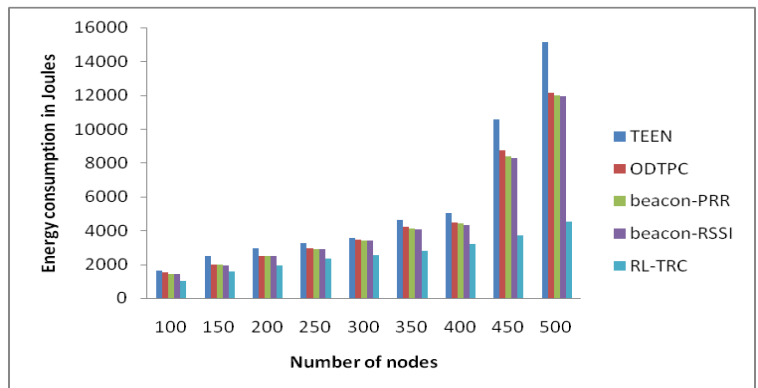
Fig 16. Energy consumption vs number of nodes (Protocol = TEEN)

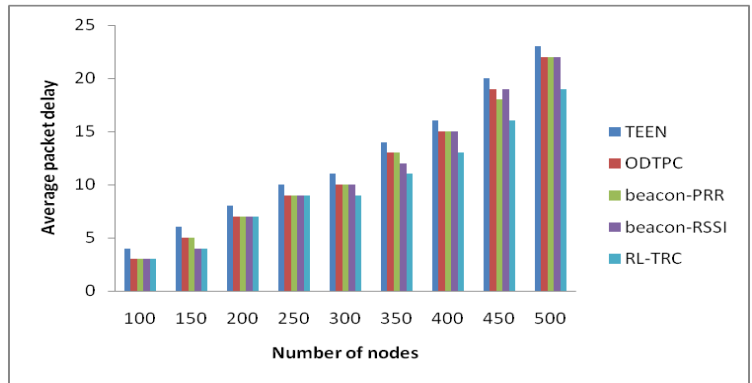
Fig 17. Average packet delay vs number of nodes (Protocol = TEEN)

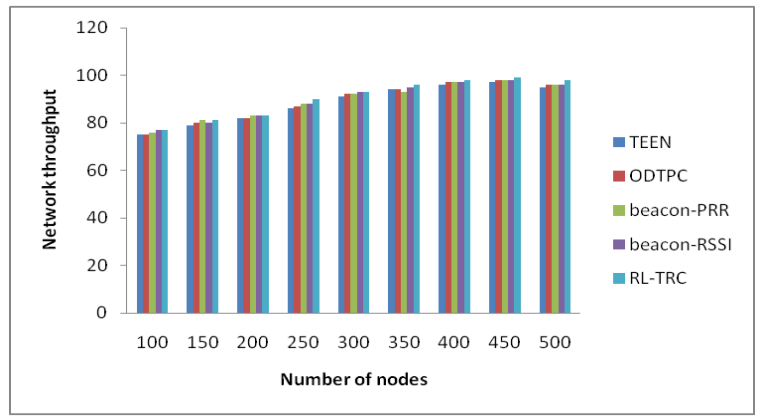
Fig 18. Network throughput vs number of nodes (Protocol = TEEN)

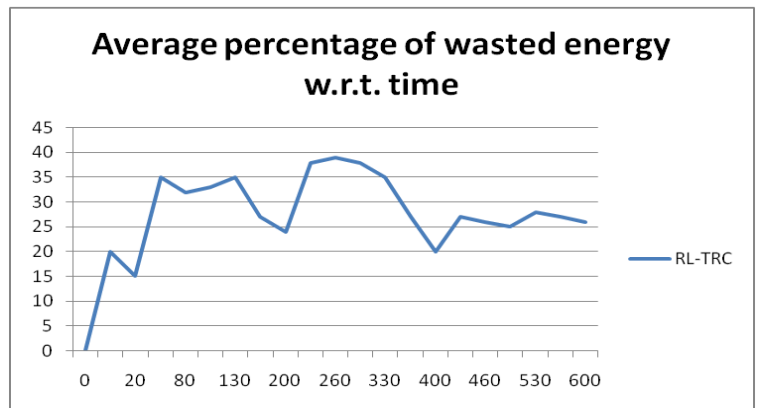
Fig 19. Average percentage of wasted energy vs Timestamp for RL-TRC

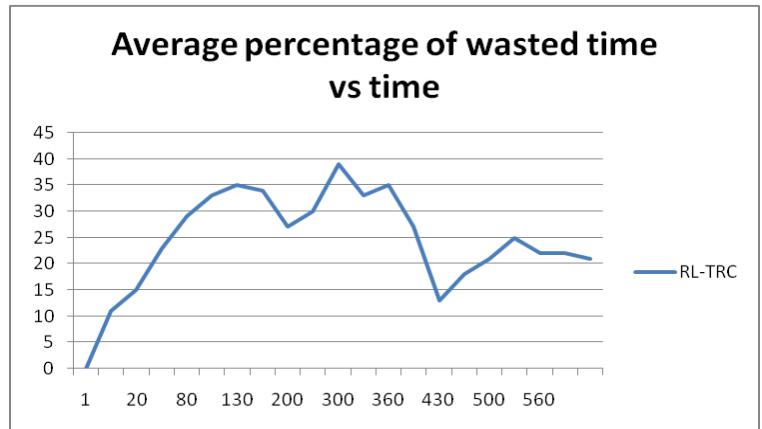

Fig 20. Average percentage of wasted time vs Timestamp for RL-TRC

Figures 19 and 20 show the average percentage of wasted energy and time with respect to timestamp. As time moves on, learning reaches its top due to an enriched number of cases. After a certain span of time, the average percentage of wasted energy and time converge due to improved experience and link quality as well as motion consciousness.

## 6. Conclusion

To the best of the authors' knowledge, the proposed transmission range control method is the first one that is suitable for motion sensors. It gains knowledge about the velocity of successors and link quality based on packet reception rate, received signal strength, and attenuation per unit distance corresponding to various power levels. Prediction of link longevity prevents transmission of signals that will never reach the desired destination. Also, the shrinking of the power level set helps nodes to identify the lower limit of the power level so that signals can survive attenuation and enable the network to avoid broadcasting of route requests for route-repair. This yields a lot of energy-saving, improving network throughput and average node lifetime, reducing delay in delivering packets from source to destination.